\documentclass[sigconf, authorversion,nonacm]{acmart}
\usepackage{multirow}
\AtBeginDocument{%
  \providecommand\BibTeX{{%
    Bib\TeX}}}

\setcopyright{acmlicensed}
\acmYear{2025}

\usepackage{hyperref}

\usepackage{tikz}
\usepackage{amsmath}

\usepackage{xcolor}
\def\BibTeX{{\rm B\kern-.05em{\sc i\kern-.025em b}\kern-.08em
    T\kern-.1667em\lower.7ex\hbox{E}\kern-.125emX}}

\usepackage{graphicx}
\usepackage{subcaption}
\usepackage[nolist]{acronym}

\usepackage[binary-units=true]{siunitx}
\usepackage{xspace}

\usepackage{todonotes}
\setuptodonotes{inline}

\newcommand{\ie}{i.\@\,e.\@\xspace}
\newcommand{\eg}{e.\@\,g.\@\xspace}

\newcommand{\etal}{et~al.\@\xspace}

\begin{document}

\title[RIS-Assisted ATR]{Anti-Tamper Radio meets Reconfigurable Intelligent Surface for System-Level Tamper Detection}

\author{Maryam Shaygan Tabar}
\affiliation{%
  \institution{Ruhr University Bochum}
  \city{Bochum}
  \country{Germany}
}
\email{maryam.shaygantabar@rub.de}

\author{Johannes Kortz}
\affiliation{%
  \institution{Max Planck Institute for Security and Privacy (MPI-SP)}
  \city{Bochum}
  \country{Germany}
}
\email{johannes.kortz@mpi-sp.org}

\author{Paul Staat}
\affiliation{%
  \institution{Max Planck Institute for Security and Privacy (MPI-SP)}
  \city{Bochum}
  \country{Germany}
}
\email{paul.staat@mpi-sp.org}

\author{Harald Elders-Boll}
\affiliation{%
  \institution{TH Köln, ICCT}
    \city{Cologne}
    \country{Germany}
}
\email{harald.elders-boll@th-koeln.de}

\author{Christof Paar}
\affiliation{%
  \institution{Max Planck Institute for Security and Privacy (MPI-SP)}
  \city{Bochum}
  \country{Germany}
}
\email{christof.paar@mpi-sp.org}

\author{Christian Zenger}
\affiliation{%
  \institution{Ruhr University Bochum,\\PHYSEC GmbH}
  \city{Bochum}
  \country{Germany}
}
\email{christian.zenger@rub.de}

\begin{abstract}

Many computing systems need to be protected against physical attacks using active tamper detection based on sensors. One technical solution is to employ an \acf{ATR} approach, analyzing the radio wave propagation effects within a protected device to detect unauthorized physical alterations. However, \ac{ATR} systems face key challenges in terms of susceptibility to signal manipulation attacks, limited reliability due to environmental noise, and regulatory constraints from wide bandwidth usage.

In this work, we propose and experimentally evaluate an \ac{ATR} system complemented by an \ac{RIS} to dynamically reconfigure the wireless propagation environment. We show that this approach can enhance resistance against signal manipulation attacks, reduce bandwidth requirements from several~GHz down to as low as 20 MHz, and improve robustness to environmental disturbances such as internal fan movements.

Our work demonstrates that \ac{RIS} integration can strengthen the \ac{ATR} performance to enhance security, sensitivity, and robustness, recognizing the potential of smart radio environments for \ac{ATR}-based tamper detection.

\end{abstract}



\maketitle

\begin{acronym}

\acro{AP}{access point}

\acro{ATR}{Anti-Tamper Radio}

\acro{AWGN}{additive white Gaussian noise}

\acro{CSI}{channel state information}

\acro{CSP}{Critical Security Parameter}

\acro{FDD}{frequency-division duplex}
\acro{FPGA}{Field Programmable Gate Array}
\acro{PUF}{Physical Unclonable Function}

\acro{HSM}{Hardware Security Module}

\acro{IP}{Intellectual property}

\acro{LoS}{line of sight}

\acro{MCS}{modulation and coding scheme}

\acro{OFDM}{orthogonal frequency division multiplexing}

\acro{PCB}{printed circuit board}

\acro{RF}{radio frequency}

\acro{RIS}{Reconfigurable Intelligent Surface}

\acro{RSSI}{received signal strength indicator}

\acro{SCA}{Side-Channel Analysis}

\acro{SNR}{signal-to-noise ratio}
\acro{JSR}{jamming-to-signal ratio}
\acro{SJNR}{signal-to-jamming-and-noise ratio}

\acro{TDD}{time-division duplex}

\acro{WLAN}{wireless local area network}
\acro{WSN}{wireless sensor network}

\acro{VNA}{vector network analyzer}

\acro{V2X}{vehicle-to-everything}

\acro{MAC}{media access control}

\acro{MIMO}{multiple-input and multiple-output}

\acro{QoS}{quality of service}

\acro{UWB}{Ultra-Wide Band}

\acro{FNR}{False Negative Rate}

\acro{FPR}{False Positive Rate}

\acro{MND}{Mean Normalized Deviation}
\end{acronym}

\section{Introduction} \label{section:Introduction}

Modern information systems constantly process and store sensitive data, including personal information, cryptographic keys, and proprietary intellectual property. The security measures to protect such data are typically designed to defend against remote attackers who intercept or manipulate communications. However, attackers with physical access fall outside the scope of such security models, as those often implicitly assume secure hardware. Therefore, if devices are not protected against physical access, attackers might succeed to compromise security. One exemplary attack scenario are supply-chain attacks, as described in the Snowden documents~\cite{spiegel_NSA} and \textit{The Big Hack} reporting~\cite{bloomberg_big2018}. The available attack tools range from side-channel analysis, fault injection, probing attacks, to Trojan insertion. To defend against such attacks, \textit{tamper detection} can be used to proactively monitor and verify physical device integrity. In this way, devices may recognize physical attacks in real-time in order to trigger defensive measures such as erasing any sensitive data. For example, tamper detection is mandatory for \acp{HSM} certified according to important standards like FIPS 140-3~\cite{fips1403} and Common Criteria~\cite{commonCriteria}.

Tamper detection solutions are designed to protect single chips or modules, employing sensor modalities that are expected to be affected by physical attacks, \eg, using switches on a device lid~\cite{immlerHigherorderAlphabetPhysical2019}, light sensors~\cite{weingartPhysicalSecurityDevices2000}, or dedicated security meshes~\cite{isaacsTamperProofTamper2013, andersonSecurityEngineeringGuide2008, obermaierPresentFuturePhysical2018}. In order to monitor the physical integrity of larger structures with complex geometry, \ie, system-level tamper detection, radio-wave propagation effects can be used. Such an \acf{ATR}~approach was recently proposed and evaluated~\cite{staatAntiTamperRadioSystemLevel2022}, enabling tamper detection for an entire 19"~server under real-world conditions. To this end, the wireless channel between a pair of antennas inside of the server is monitored for changes, allowing the detection of minute physical perturbations such as the insertion of sub-mm diameter probing needles. The advantages of this approach are the inherent flexibility of wireless radio propagation, the possibility of retrofitting existing devices, and re-initialization. However, several open challenges remain for the \ac{ATR} approach. In this paper, we study programmable radio enclosures based on \acp{RIS} to address the following three key challenges (C1~-~C3) of \ac{ATR} systems:
\begin{itemize}
\setlength{\itemsep}{2pt}
    \item \textbf{Security Against Signal Manipulation (C1):} 
    At its core, the \ac{ATR} employs reception of radio signals to sense tamper events in the environment. However, due to the open nature of the wireless medium, an attacker could attempt manipulation of the sensing result to evade tamper detection. The \ac{ATR} proposal~\cite{staatAntiTamperRadioSystemLevel2022} did not take such attacks into account, assuming that the attacker cannot anticipate the effect of their tampering. This reasoning is rooted in the complexity of the physical environment, making it practically infeasible for an attacker to obtain the channel response. However, we argue that a determined attacker might attempt to replicate or simulate the environment to assess how the \ac{ATR} behaves under a physical attack.
        
    \item \textbf{Usability Optimization (C2):}
    A precondition to detect a tamper event is broad signal coverage inside the enclosure. This ensures that the attacker interacts with the radio waves when tampering.
    However, when integrating an \ac{ATR} with an existing computing system, it is difficult to anticipate signal coverage, depending on antenna placement and electromagnetic properties of the environment. Moreover, the environment might exhibit varying radio propagation even without tampering, \eg, due to moving fans on the inside of the enclosure.

    \item\textbf{Spectrum Usage (C3):}
    The \ac{ATR} proposal~\cite{staatAntiTamperRadioSystemLevel2022} partially tackled the previous challenge~C2 through wideband signals, exploiting frequency diversity to enhance average spatial signal coverage and while allowing to exclude noisy parts of the signal. However, this approach allocates a total of \SI{7}{GHz} bandwidth which might yield compliance and regulatory issues. In particular, such a signal design might face spectrum allocation constraints, potentially causing and experiencing interference with regulated frequency bands.

\end{itemize}

In this work, we resolve these challenges by exploiting \ac{RIS}-controlled adjustment of the radio propagation within the \ac{ATR}-protected environment. In particular, we resolve C1 by making the radio environment unpredictable for an attacker. Moreover, resolving C2, we show that the \ac{RIS} can adapt the environment to improve the \ac{ATR} performance in terms of robustness against environmental noise and detection sensitivity.  Finally, we demonstrate that this even works when making the \ac{ATR} bandwidth as small as~\SI{20}{MHz}, resolving C3.

In this paper, we make the following key contributions:
\begin{itemize}
\setlength{\itemsep}{2pt}
    \item We are the first to propose and study the combination of \ac{ATR}-based tamper detection and the \ac{RIS} technology.

    \item We discuss and analyze the possibility of signal manipulation attacks against an \ac{ATR} system and show that an \ac{RIS}-assisted \ac{ATR} can resist such an attack. In particular, we randomize the radio environment, linking \ac{ATR} security to the unpredictability of the radio environment.

    \item We show that a \ac{RIS}-programmable radio environment can be tuned towards improved \ac{ATR} tamper detection performance in terms of detection robustness and sensitivity. In this novel design approach, we jointly tune the \ac{ATR} measurement system and the monitored environment. 
\end{itemize}

\section{Background}

In this section, we provide technical background on physical tamperproofing, the \ac{ATR} approach, and the \ac{RIS} technology.

\subsection{Tamperproof Hardware}
Physical tamperproofing falls into three categories:
($i$) \textit{Tamper evidence} ensures tampering leaves detectable traces for after-the-fact inspection, using irreversible materials like seals or coatings~\cite{appelSecuritySealsVoting2011, andersonSecurityEngineeringGuide2008}. 
($ii$) \textit{Tamper resistance} makes physical access difficult, increasing attack effort and cost, as seen in bank safes or epoxy-coated \acp{PCB}~\cite{weingartPhysicalSecurityDevices2000}.
($iii$) \textit{Tamper detection} actively monitors for tamper events via sensor inputs like vibration, light, or radio-wave propagation~\cite{weingartPhysicalSecurityDevices2000, staatAntiTamperRadioSystemLevel2022}. Detection triggers a tamper response, typically wiping sensitive data, \eg, cryptographic keys or \ac{IP}~\cite{grandPracticalSecureHardware2004, dattaTechniquesTamperDetection2022}.

\paragraph{Tamper Detection System Design Goals}
In the scheme by Immler~\cite{immlerHigherorderAlphabetPhysical2019}, tamper detection balances the following goals:

($i$) \textit{Security}: The primary system goal is to detect against any tamper event with a true-positive rate as high as possible. To resist attacks on itself, the detection mechanism should be part of the protection scope. ($ii$) \textit{Usability}: This requirement calls for a low false-positive rate, \ie, expected legitimate conditions should not trigger the tamper response, which would render the device unusable despite the absence of an attack. ($iii$) \textit{Manufacturing}: Manufacturing of an anti-tamper solution should be feasible with moderate technological effort regarding skill and tooling. The key factors here are cost and sourcing issues. The intricate manufacturing processes of~\acp{HSM} is described by Isaacs~\etal~\cite{isaacsTamperProofTamper2013}.

\subsection{\acf{ATR}}
\label{subsec:atr_background}

Staat~\etal~\cite{staatAntiTamperRadioSystemLevel2022} proposed the \ac{ATR}, employing a wireless sensing approach to realize system-level tamper detection. The \ac{ATR} idea is based on the observation that wireless radio propagation effects reflect the physical state of an environment, enabling detection of minute physical perturbations in computing systems. To verify physical integrity, the \ac{ATR} permanently monitors a \SI{7}{GHz} bandwidth channel frequency response between two antennas on the inside of a protected environment and compares it to an initial reference measurement.

The lifecycle of an \ac{ATR} system encompasses two phases: 

\paragraph{Provisioning}
The \ac{ATR} is provisioned at time $t_0$ in a trusted environment to record an initial reference measurement that is known to reflect the physical state of the environment in absence of an attack. We denote this initial reference measurement as $H_{R}(f, t_0)$. Further, additional training steps may be performed, \eg, to identify and discard unreliable parts of the \ac{ATR} channel response.

\paragraph{Deployment} 
After provisioning, the \ac{ATR} channel measurements are repeated at times $t_i$ where $i > 0$, observing a response that deviates by $\Delta H(f,t_i)$ from the initial reference channel $H_{R}(f, t_0)$. This drift component can be caused by legitimate environmental variation due to variation in temperature, humidity, and mechanical changes such as vibration or fan movement. As the \ac{ATR}-protected device is deployed in a potentially adversarial environment, an attacker might attempt to gain physical access, causing the attacker imprint $\Delta H^{(T)}(f, t_i)$. Additionally, the measurement is affected by \ac{AWGN} $N$.
\begin{equation}
    H(f, t_i) = H_{R}(f, t_0) + \Delta H(f,t_i) + \Delta H^{(T)}(f, t_i) + N.
    \label{eq:atr_deployed}
\end{equation}
While each contribution denotes a complex-valued channel gain, channel magnitude is typically used to simplify measurement. In order to verify physical integrity, the \ac{ATR} compares occurring measurements to the initial reference using a distance metric, \ie, $dist(H_{R}(f, t_0), H(f, t_i))$. A tamper event is declared if the metric exceeds a threshold $\mathcal{T}$. Ideally, the distance metric exhibits sensitivity to the attacker imprint $\Delta H^T$ while being robust against other deviations.

\subsection{\acf{RIS}}

An \ac{RIS} is an engineered surface to digitally control reflections of radio waves, enabling \textit{smart radio environments}, \eg, for future 6G networks~\cite{jiangRoad6GComprehensive2021}. Here, the radio propagation medium is considered as a degree of freedom to optimize wireless communication by redirecting radio waves in certain directions~\cite{liaskosNovelCommunicationParadigm2019}, \eg, to improve signal coverage and eliminate dead zones. An \ac{RIS} passively reflects existing ambient signals by utilizing $L$~identical unit-cell reflector elements arranged on a planar surface. Importantly, a programmable \ac{RIS} configuration vector denoted as $c$ allows individually controlling the reflection coefficient of each reflector, \eg, to select between two reflection phases \SI{0}{\degree} and \SI{180}{\degree}, corresponding to the reflection coefficients~$+1$ and~$-1$. Typically, an \ac{RIS} is realized as a \ac{PCB} with printed microstrip reflectors, enabling very low-cost implementation~\cite{ranaReviewPaperHardware2023}, using $L$~digital logic signals, \eg, from a microcontroller to represent the configuration $c$.

\section{\acs{RIS}-assisted \acs{ATR}}

In this section, we outline the system model of an \ac{RIS}-assisted \ac{ATR}, introduce the attacker model, and describe our experimental setup. 

\subsection{System Model And Attacker}

\begin{figure}[htb]
    \centering
    \includegraphics[width=\linewidth]{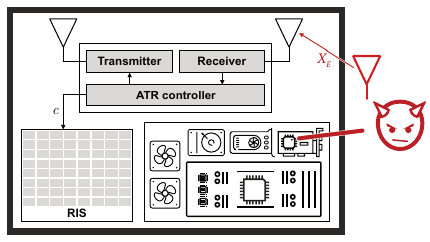}
    \caption{Illustration of an \ac{RIS}-assisted \ac{ATR} system, including an attacker attempting a physical perturbation and being capable of signal injection.}
    \label{fig:ris_atr_system}
\end{figure}

In this work, we consider a computing system that utilizes an \ac{ATR} as described in~\autoref{subsec:atr_background} to detect physical tampering, defined by violation of the physical integrity of the device. We assume that the radio propagation within the enclosure is partially controllable by means of an \ac{RIS} which is controlled by the \ac{ATR} subsystem, as illustrated in~\autoref{fig:ris_atr_system}. We extend the initial \ac{ATR} provisioning phase to collect a set of reference measurements, each corresponding to an individual \ac{RIS} configuration. To take the \ac{RIS} effect on the device-internal wireless channel into account, we adjust~\autoref{eq:atr_deployed} as follows, incorporating an additional dependency on the \ac{RIS} configuration $c_i$:
\begin{equation}
    H(f,t_i,c_i) = H_R(f,t_0, c_i) + \Delta H(f,t_i, c_i) + \Delta H^{(T)}(f,t_i,c_i) + N. \label{eq:atr_system_eq}
\end{equation}

\paragraph{Attacker Model}
The attacker's goal is to physically perturb the environment without triggering a detection by the \ac{ATR} system. In the context of this work, we assume an attacker inserting a thin metallic needle into the environment, causing the \ac{ATR} channel response to change by $\Delta H^{(T)}(f,t_i,c_i)$, assuming that such subtle tampering subsumes other more invasive attack strategies. Other than in the initial \ac{ATR} proposal~\cite{staatAntiTamperRadioSystemLevel2022}, we assume that the attacker is aware of the reference channel $H_{R}(f,t_0)$ and the effect of their physical attack, $\Delta H^{(T)}(f,t_i,c_i)$. The attacker might obtain such information by means of cloning or simulating the environment. Crucially, however, we assume the attacker does not know the \ac{RIS} configuration~$c_i$. Moreover, we grant the attacker the capability of injecting a radio signal~$X_E$ into the the \ac{ATR} receiver, \eg, to manipulate the \ac{ATR} channel measurement.

\subsection{Experimental Setup}

We now describe our experimental setup, including the \ac{RIS}-controlled protected environment, the automated probing attack testbed, and the radio measurements.

\paragraph{Attack Simulation}

For our experimental setup, we replicate the methodology introduced by Staat~\etal~\cite{staatAntiTamperRadioSystemLevel2022}, using a a three-axis Cartesian robot to automatically and repeatably insert probing needle into a test enclosure. We mount a metallic probing needle with a diameter of \SI{0.3}{\mm} and length \SI{4}{\cm} on a slide on the z-axis. Thus, we use the robot to control the position and depth of the needle insertion into the test enclosure.

\begin{figure}
    \centering
    \includegraphics[width=1.0\linewidth]{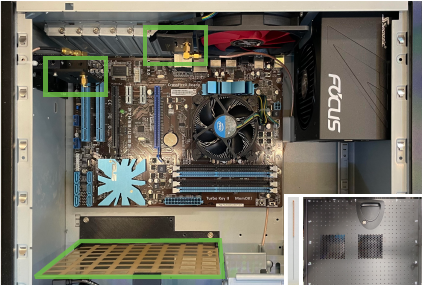}
    \caption{Inside of the computer enclosure with two \ac{ATR} antennas and the \ac{RIS}. The inset shows the top lid with pre-made holes for needle insertion and and a close-view of the \SI{0.3}{mm} diameter probing needle.}
    \label{fig:experimental_setup}
\end{figure}

\paragraph{Computer Enclosure}
We investigate the \ac{RIS}-assisted \ac{ATR} using a desktop computer with dimensions 43~$\times$~43~$\times$~\SI{20.5}{\cm}. The enclosure houses a functional computer with a motherboard and peripherals, including fans, PCI cards, and a power supply. Additionally, we used 3D-printed parts to install two taoglas~FXUWB10 wideband patch antennas and an \ac{RIS}.

We drilled holes in a grid pattern into the top lid, allowing insertion of probing needles to to simulate a reproducible tamper events. Due to the sub-wavelength diameter, the holes have negligible effect on the \ac{RF} shielding of the case. In our experiments, we performed probing at a total of 250 positions.

\paragraph{\acs{RIS}}

We added an \ac{RIS} to control the radio signal propagation within the computer case. It is based on the open-source design of (removed for anonymity)~\etal~[X, X] 
and is built as a standard FR4 \ac{PCB}. It has $64$ unit-cell reflector elements with binary phase control, where each reflection coefficient can either have phase \SI{0}{\degree} (state~`\texttt{0}') or \SI{180}{\degree} (state~`\texttt{1}') in the \SI{5}{\GHz} frequency range. The \ac{RIS} is controlled via USB to program the 64-bit configuration word $c$. See~[X] %
for additional technical details.

To find \ac{RIS} configurations which yield certain desired channel properties, \eg, minimization of the channel magnitude and the standard deviation, we employed a greedy genetic \ac{RIS} optimization algorithm from the literature~\cite{tewesIRSenabledBreathTracking2022}. The algorithm evaluates the \ac{RIS}-element-wise empirical probability for minimizing a cost function to generate new configuration candidates that are increasingly likely to meet the optimization goal.

\paragraph{Radio Channel Measurements}
We use a Keysight P9372A~\ac{VNA} to collect complex-valued frequency responses of the channel between the \ac{ATR} antennas~\cite{p9372a_manual}. For the measurements, we use an IF bandwidth of \SI{10}{\kHz} to sample the frequency response in \SI{1}{\MHz} steps. For experimental simplicity, we place the \ac{VNA} outside of the computer case.

\paragraph{Detection Metrics}
To distinguish \ac{ATR} measurements corresponding to the untampered and tampered states of the protected environment, we employ the \ac{MND} and Euclidean distance metrics. With these metrics, we quantify deviations from an initial reference measurement. According to~\cite{staatAntiTamperRadioSystemLevel2022}, we employ the \ac{MND} to compare wideband measurements while we resort to the Euclidean distance to compare narrowband measurements.

\paragraph{Threshold Selection}
In our experiments, we use an initial untampered measurement as the reference $H_R(f, t_0)$ against which we compare all subsequent measurements using the respective distance metric. To evaluate the tamper detection performance at a false-alarm rate of 0, we select the detection threshold as the maximum observed distance in the untampered state. Thus, we can indicate the \ac{FNR} to quantify the detection sensitivity.

\section{Experiments And Results}

In this section, we experimentally study the utility of the \ac{RIS} to complement an \ac{ATR} system. To this end, we resolve the challenges C1-C3 as mentioned in the Introduction.

\subsection{Bandwidth Reduction (C3)}

To achieve wide detection coverage, the initial \ac{ATR} system proposal~\cite{staatAntiTamperRadioSystemLevel2022} employs wideband channel responses, \eg, covering \SI{7}{\GHz} bandwidth. However, such extensive spectral allocation can be undesired due to regulatory requirements and radio co-existence. Reducing the high bandwidth requirement of an \ac{ATR} system corresponds to challenge C3 outlined in the Introduction. In the following, we investigate the impact of bandwidth reduction on the \ac{ATR} performance and how to employ the \ac{RIS} in this context.

\subsubsection{Tamper Detection with Narrowband Measurements}
In our first experiment, we study how the signal bandwidth affects the detection of probing needle insertions into the computer case. For this, we collect \SI{7}{\GHz} bandwidth frequency responses with 7000 frequency points, allowing us to extract narrowband segments to evaluate the needle detection performance on. In our experiment, we move the robot to each of the 250~drill positions and measure the frequency response with the needle positioned outside (untampered) and inserted \SI{20}{\mm} deep into the respective drill hole (tampered). 

\begin{figure}[htb]            
    \centering
    \begin{subfigure}[b]{0.225\textwidth}
        \centering
        \includegraphics[width=1\linewidth]{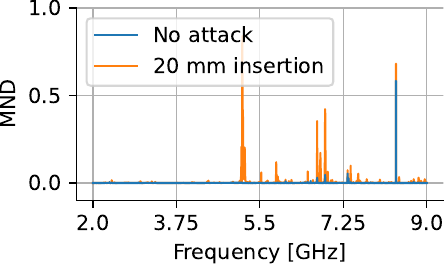}
        \caption{Drill position A}
    \end{subfigure}
    \quad
    \begin{subfigure}[b]{0.225\textwidth}
        \centering
        \includegraphics[width=1\linewidth]{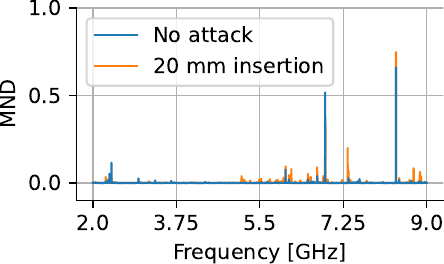}
        \caption{Drill position B}
    \end{subfigure}
    \quad
    
    \begin{subfigure}[b]{0.225\textwidth}
        \centering
        \includegraphics[width=1\linewidth]{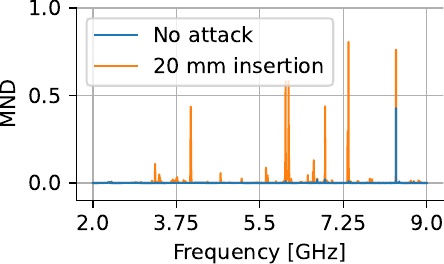}
        \caption{Drill position C}
    \end{subfigure}
    \quad
    \begin{subfigure}[b]{0.225\textwidth}
        \centering
        \includegraphics[width=\linewidth]{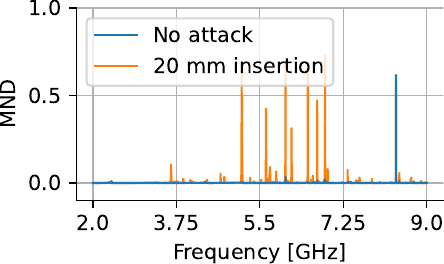}
        \caption{Drill position D}
    \end{subfigure}
\caption{Effect of a \SI{20}{mm} needle insertion over frequency for four different needle insertion positions, each being unique to to the position.}
\label{fig:detection_performance_across_freq}
\end{figure}

We found that the needle insertion affects the \ac{ATR} channel response differently for each needle position. This effect can be seen from \autoref{fig:detection_performance_across_freq}, showing the effect of the needle insertion on the frequency response for four different drill hole positions. Here, we observe narrowband deviations at different frequencies, depending on the respective drill position. Thus, when reducing bandwidth, it becomes less likely to capture these deviations -- potentially causing a false-negative detection.

\begin{figure}[htb]
\centering
\begin{subfigure}[b]{0.4\textwidth}
    \centering
    \includegraphics[width=1\linewidth]{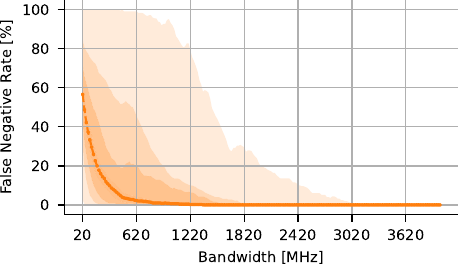}
    \caption{}
    \Description{todo}
    
\end{subfigure}
\quad
\begin{subfigure}[b]{0.4\textwidth}
    \centering
    \includegraphics[width=1\linewidth]{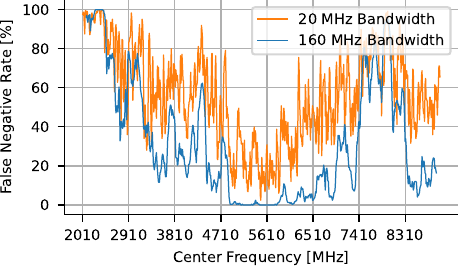}
    \caption{}
    \Description{}
\end{subfigure}
\caption{(a) \ac{FNR} across various bandwidths. The shaded area represents the variation in \ac{FNR} for different center frequencies. (b) \ac{FNR} across various center frequencies.}
\label{fig:bandwidth_vs_dr.}
\end{figure}

\begin{figure}[htb]
           \centering
           \begin{subfigure}[b]{0.225\textwidth}
               \centering
               \includegraphics[width=1\linewidth]{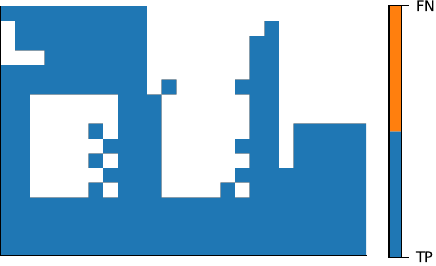}
               \caption{$f_c=\SI{5.5}{GHz}$ and \SI{7}{GHz} Bandwidth}
               \Description{todo}
               \label{}
           \end{subfigure}
           \quad
           \begin{subfigure}[b]{0.225\textwidth}
               \centering
               \includegraphics[width=1\linewidth]{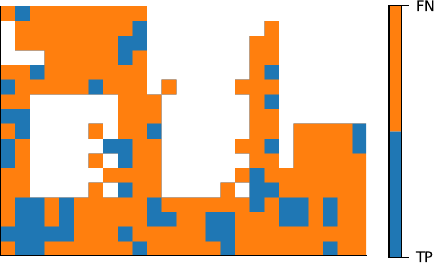}
               \caption{$f_c=\SI{4.79}{GHz}$ and \SI{20}{MHz} Bandwidth}
               \Description{todo}
               \label{}
           \end{subfigure}
       \caption{The impact of bandwidth reduction on detection: (a) With a 7 GHz bandwidth, the \ac{FNR} is zero. (b) With a 20 MHz bandwidth, the \ac{FNR} is $77.2\%$.}
       \label{fig:heatmap_bandwidth_reduction}
       \end{figure}

\autoref{fig:bandwidth_vs_dr.} shows the impact of different bandwidths and center frequencies on false negative rates. Here, we can see that the false negative rate increases as we reduce the bandwidth. However, this effect depends much on the particular center frequency being used. Overall, we observe that the needle detection works best between approx.~\SI{4.7}{\GHz} and~\SI{6.5}{\GHz}, yet with significant variations. For example, when using $f_c=\SI{4.79}{GHz}$ with 20~MHz bandwidth, the \ac{FNR} drops to \SI{77.2}{\percent}, as shown by  \autoref{fig:heatmap_bandwidth_reduction}, which depicts spatial detection.

\subsubsection{\ac{RIS} Optimization for Enhanced Detection}
\begin{figure}[htb]
    \centering
    \includegraphics[width=0.85\linewidth]{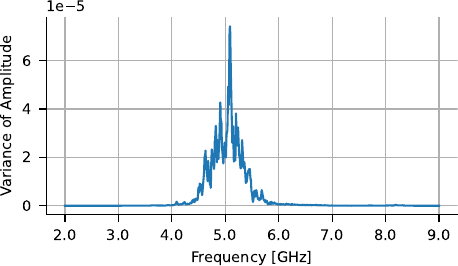}
    \caption{Variation across 100 random \ac{RIS} configurations to identify the region where \ac{RIS} has the most effect.}
    \Description{\todo{caption and description}}
    \label{fig:ris_vs_frq}
\end{figure}

In the previous experiment, the \ac{RIS} on the inside of the computer was set to a fixed configuration. However, the \ac{RIS} allows us to partially reconfigure the radio propagation within the computer case which promises to improve the \ac{ATR} performance. First of all, we investigate how the \ac{RIS} affects the \ac{ATR} channel response. For this, we measure the response of 100 random \ac{RIS} configurations and calculate the variance of the channel across the \ac{RIS} configuration as shown in \autoref{fig:ris_vs_frq} as a function of frequency. Here, we can see that the \ac{RIS} effect on the channel is strongest between approx.~\SI{4.5}{\GHz} and ~\SI{5.5}{\GHz}.

\begin{figure}
        \centering
        \includegraphics[width=.75\linewidth]{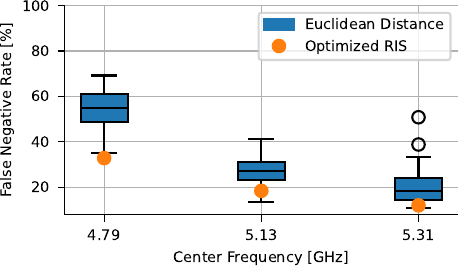}
        \caption{\ac{FNR} across three different center frequencies and a \SI{20}{MHz} bandwidth, evaluated with 50 random \ac{RIS} configurations and an optimized \ac{RIS} configuration.}
        \label{fig:boxplot_fnr}
    \end{figure}

As we will show next, the needle detection performance within the effecitve \ac{RIS} frequency range depends on the particular \ac{RIS} configuration. We study this effect by evaluating the needle detection within a \SI{20}{MHz} bandwidth for the center frequencies \SI{4.79}{GHz}, \SI{5.13}{GHz}, and \SI{5.31}{GHz}. 
For the experiment, we first generate 50~random \ac{RIS} configurations. We then run the same procedure as before, measuring the channel responses with the needle outside and \SI{20}{\mm} deep inside the computer case. This time, however, for each needle position, we cycle through the 50~\ac{RIS} configurations and collect 50~corresponding frequency responses. \autoref{fig:boxplot_fnr} depicts the distribution of the \acp{FNR} accross \ac{RIS} configurations, demonstrating that the \ac{RIS} can both improve and degrade the detection. For example, at \SI{4.79}{GHz}, the \ac{FNR} varies between \SI{36}{\percent} and \SI{70}{\percent}.
        
\begin{figure}[htb]
       \centering
       \begin{subfigure}[b]{0.225\textwidth}
           \centering
           \includegraphics[width=1\linewidth]{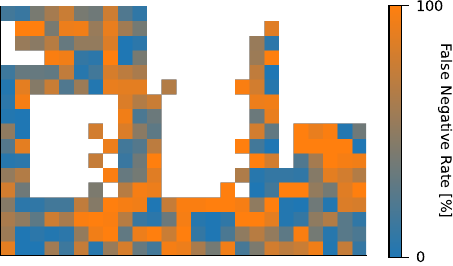}
           \caption{Random $c_i$}
           \label{}
       \end{subfigure}
       \quad
       \begin{subfigure}[b]{0.225\textwidth}
           \centering
           \includegraphics[width=1\linewidth]{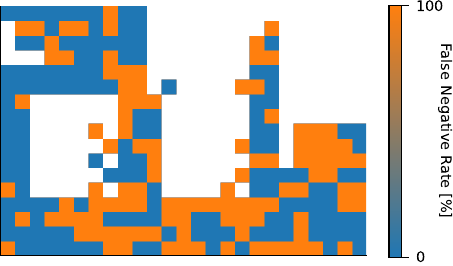}
           \caption{Optimized $c_i$}
       \end{subfigure}
   \caption{The impact of \ac{RIS} optimization of the spatial distribution of \ac{FNR} with $f_c=\SI{4.79}{GHz}$ and \SI{20}{\MHz} bandwidth. (a) Random configurations $c_i$. (b) \ac{RIS} is optimized to minimize the average channel amplitude.}
   \label{fig:detection_improvement}
   \end{figure}

The previous result underlines the potential of the \ac{RIS} to control the tamper detection sensitivity. However, in order to leverage this constructively for the \ac{ATR}, we need to find a well-suited \ac{RIS} configuration that improves and not degrades the detection performance. This this is trivial in the setting where we can test each \ac{RIS} configuration against the attacker effect. However, in an actual deployment scenario, the key challenge rather is to find well-suited \ac{RIS} configurations \textit{blindly}, \ie, without any knowledge of the attacker effect. In our experiments, we found that we can resolve this challenge by using the \ac{RIS} to \textit{minimize} the channel amplitude, \ie, without any knowledge about the attacker. In our experiments, we optimized the \ac{RIS} to minimize the amplitudes of a \SI{20}{\MHz} channel at three center frequencies \SI{4.79}{GHz}, \SI{5.13}{GHz}, and \SI{5.31}{GHz} and then employed the resulting configurations to repeat the needle probing. The results are shown by the red markers in \autoref{fig:boxplot_fnr}, indicating the \ac{FNR} of the optimized \ac{RIS}. Here, we can see that the optimized \ac{RIS} configurations significantly reduces the \ac{FNR} detection performance compared to the randomly configured \ac{RIS}, ranking in the lower end of the distribution. \autoref{fig:detection_improvement} shows this effect in the spatial domain, indicating the average \ac{FNR} for each drill hole position. Thus, we can conclude that minimizing the channel amplitude enhances the \ac{ATR} detection sensitivity. That is because the minimization demands many propagation paths to destructively interfere with each other, being very sensitive against subtle signal changes, \eg, caused by inserting the probing needle.

\subsection{Security Hardening (C1)}

Based on the previous insight that the \ac{RIS} makes the channel response on the inside of the computer case programmable, we will next show that this allows enhancing the security of the \ac{ATR} against signal manipulation attacks. We first explain how the attack works and then outline how the \ac{RIS} can be used to thwart it. %

\subsubsection{Attack Scenario}

We assume that the attacker knows the effect of their physical tampering on the \ac{ATR} channel response, \eg, by exactly recreating the targeted environment. In view of the wireless nature of the \ac{ATR}, the attacker can then perform a signal injection attack to manipulate the \ac{ATR} sensing system to remove the effects of the physical tampering $\Delta H^T(f, t_{i})$. In particular, the attacker sends a signal $X_E$ towards the \ac{ATR} receiver that causes the additive effect $\Delta H_A^T(f, t_{i}) \approx -\Delta H^T(f, t_{i})$ on the channel measurement. In practice, the attacker would also need to account for the channel between their antenna and the \ac{ATR} receiver. However, in the following, we pessimistically assume a direct injection of the compensation signal into the \ac{ATR} receiver, \ie, a unity channel response. %
The compensated system equation then is as follows:
\begin{equation}\label{eq:double}
    \begin{aligned}
        \tilde H(f,t_i) =\ & H_{R}(f,t_0) + \Delta H(f,t_i) \\ 
        & + \Delta H^{(T)}(f,t_i) + \Delta H_A^T(f, t_{i}) + N \\
        & \approx H_{R}(f,t_0) + \Delta H(f,t_i) + N.
    \end{aligned}
\end{equation}\label{eq:attack}
What remains after cancellation approximates the \ac{ATR} channel response in absence of a tamper event, deviating from the initial reference only by the legitimate environmental drift $\Delta H(f,t_i)$. Thus, the \ac{ATR} is likely to miss the needle detection, \ie, a false negative detection.

\subsubsection{\ac{RIS}-based Defense Strategy}
We assume that the attacker does not know the \ac{RIS} configuration~$c_i$ and therefore cannot exactly anticipate the propagation effects on the inside of the computer case. Crucially, when the attacker imprint $\Delta H^T(f, t_{i}, c_i)$ now is a function of the \ac{RIS} configuration, this allows us to prevent the attacker from crafting the correct compensation signal.

\begin{figure}
    \centering
    \includegraphics[width=1.0\linewidth]{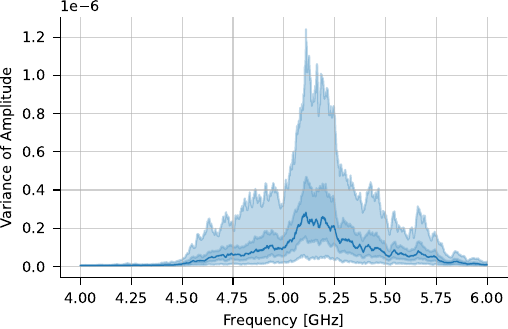}
    \caption{\ac{RIS}-introduced variance of amplitude of $\Delta H^T$. The median as well as the distribution over all needle positions can be seen.}
    \label{fig:var_attacker_effect}
\end{figure}

First, we evaluate the \ac{RIS} impact on  $\Delta H^T$. For this, we measure 100 random \ac{RIS} configurations $c_i$ over a \SI{7}{GHz} bandwidth. For each of the 250 drill holes and each \ac{RIS} configuration, we measure the channel responses without and with tampering (\SI{20}{\mm} needle insertion). To estimate the attacker imprint $\Delta H^T(f, t_i, c_i)$ we subtract the complex-valued channel responses corresponding to the tampered and untampered environment. Then we calculate the variance of the amplitude of $\Delta H^T(f, t_i, c_i)$ over the \ac{RIS} configurations $c_i$. \autoref{fig:var_attacker_effect} shows the distribution of the \ac{RIS}-induced variance on $\Delta H^T$ across drill positions over frequency. The important observation here is that $\Delta H^T$ is indeed affected by the \ac{RIS} configuration~$c_i$. As expected, the frequency-dependence of this effect lines up with~\autoref{fig:ris_vs_frq}.

With the \ac{RIS} in place, the system equation for a compensated tamper event changes as follows:
\begin{equation}
    \begin{aligned}
        \tilde H(f,t_i, c_i) =\ & H_{R}(f,t_0, c_i) + \Delta H(f,t_i, c_i)\\
        & + \Delta H^{(T)}(f,t_i, c_i) + \Delta H_A^T(f, t_{i'}, c'_{i}) + N.
    \end{aligned}
\end{equation}\label{eq:attack_with_ris}
Thus, the attacker is now forced to craft their compensation signal in order to match the \ac{RIS} configuration, \ie, $c_i = c'_{i}$. Since the attacker does not know $c_i$, they might attempt to randomly guess $c'_{i}$. Another strategy would be to construct an average compensation signal. Here, we assume the attacker can observe $\Delta H^T$ for different \ac{RIS} configurations within their replica environment and then use the average of $\Delta H^T$ over $c'_{i}$.

\begin{figure}
\centering
\begin{subfigure}[b]{0.225\textwidth}
    \includegraphics[width=1.0\linewidth]{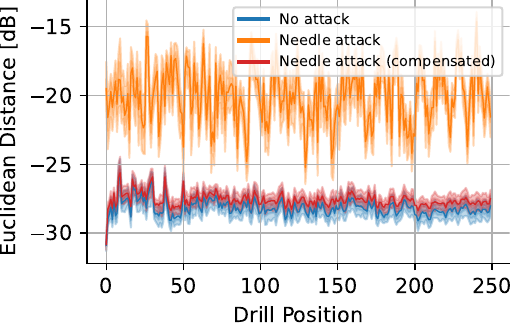}
    \caption{Without \ac{RIS}.}
    \label{fig:10a}
\end{subfigure}
\quad
\begin{subfigure}[b]{0.225\textwidth}
    \includegraphics[width=1.0\linewidth]{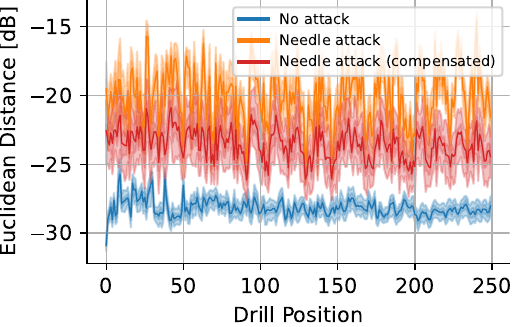}
    \caption{With \ac{RIS}.}
    \label{fig:10b}
\end{subfigure}
\begin{subfigure}[b]{0.225\textwidth}
    \includegraphics[width=1.0\linewidth]{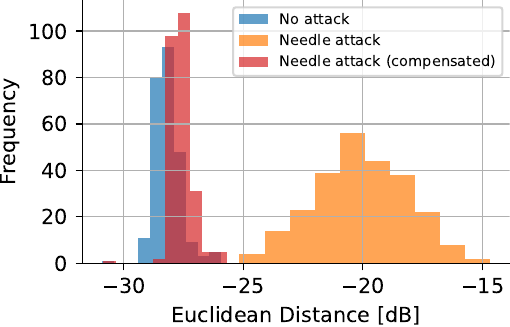}
    \caption{Without \ac{RIS}.}
    \label{fig:10c}
\end{subfigure}
\quad
\begin{subfigure}[b]{0.225\textwidth}
    \includegraphics[width=1.0\linewidth]{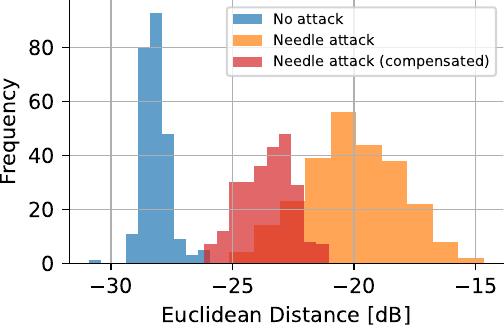}
    \caption{With \ac{RIS}.}
    \label{fig:10d}
\end{subfigure}
\caption{Euclidean distance in the compensation attack scenario with \SI{160}{MHz} channel bandwidth. Distributions are calculated over different random \ac{RIS} configurations. (a) and (c) show a successful attack, where the attacker possesses the correct \ac{RIS} configurations. (b) and (d) show the performance when the attacker employs an average compensation signal.}
\label{fig:attack_distances}
\end{figure}

In our first experiment, we evaluate the attack scenario for a channel bandwidth of \SI{160}{\MHz}, corresponding to a common channel bandwidth of, \eg, \mbox{Wi-Fi} systems. \autoref{fig:10a} and~\autoref{fig:10c} depict the effect of the attacker's signal injection attack designed to remove the effect of the tampering. Here, we plot the Euclidean distances of the \ac{ATR} measurements to the initial reference measurement. The key observation is that the attacker succeeds to reduce the Euclidean distance for each drill position significantly towards the legitimate system variation, likely evading tamper detection and increasing the \ac{FNR}. In contrast, \autoref{fig:10b} and~\autoref{fig:10d} show the attack performance against the \ac{RIS}-assisted \ac{ATR} where the attacker does not possess the correct $c_i$. Instead, the attacker resorts to injecting an average of $\Delta H^T$. We observe that this attack strategy allows to the attacker to reduce the impact of their tampering. Importantly, however, the Euclidean distances with and without an attack remain separable.

\begin{figure}[h]
\centering
\begin{subfigure}[b]{0.225\textwidth}
    \includegraphics[width=1.0\linewidth]{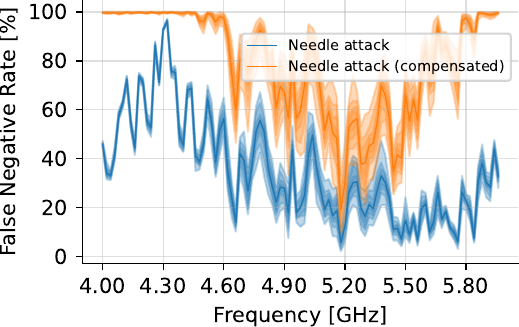}
    \caption{Bandwidth: \SI{20}{MHz}}
    \label{fig:12a}
\end{subfigure}
\quad
\begin{subfigure}[b]{0.225\textwidth}
    \includegraphics[width=1.0\linewidth]{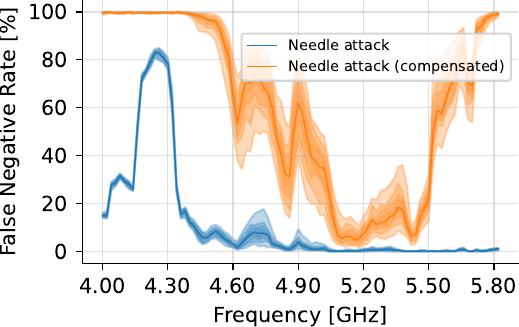}
    \caption{Bandwidth: \SI{160}{MHz}}
    \label{fig:12b}
\end{subfigure}
\caption{Relation between center frequency and \ac{FNR} of the average compensation attack. (a) With \SI{20}{MHz} bandwidth and (b) with \SI{160}{MHz}.}
\label{fig:attack_compensation_counter}
\end{figure}

Since we have already seen strong variation in the effectiveness of the \ac{RIS} at different center frequencies, we now look at how this affects the attacker's success rate. In \autoref{fig:attack_compensation_counter} we look at the \ac{FNR} of the attacker with compensation and \ac{RIS} countermeasure. The frequencies in \autoref{fig:12a} are measured with \SI{20}{MHz} bandwidth and in \autoref{fig:12b} with \SI{160}{MHz}.  

An inverse proportionality to \autoref{fig:ris_vs_frq} can be seen for both bandwidths. In the frequency ranges where the \ac{RIS} generates the most variation, the attack is the least successful. Regardless of the bandwidth, the \ac{FNR} can drop to below \SI{10}{\%} at maximum \ac{RIS} variance. This means that in addition to security in a reinforced attacker model, the \ac{RIS} also simplifies the usability of the system at the same time, as the variance of the \ac{RIS} provides us with a simple metric for selecting a suitable frequency with a low bandwidth.

\begin{figure}[h]
    \centering    \includegraphics[width=0.75\linewidth]{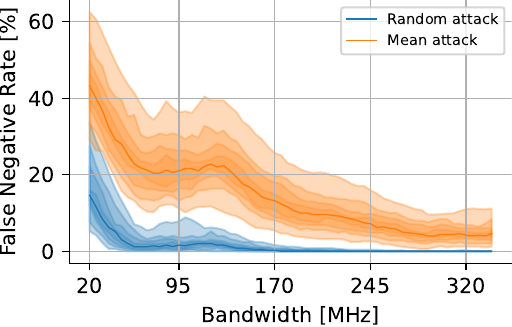}
    \caption{Compensation attacker success in terms of the \ac{FNR} against the \ac{RIS}-assisted \ac{ATR} over channel bandwidth. %
    }
    \label{fig:attack_vs_bw}
\end{figure}

Now we look at how the attacks behave with different bandwidths. The measurement data is generated in the same way as before, using a center frequency of \SI{5.31}{GHz}. In \autoref{fig:attack_vs_bw}, both attack methods are shown, with the distribution over reference configurations shaded and the median highlighted in bold. As expected, higher bandwidths yield lower \ac{FNR} values. Notably, a bandwidth of \SI{320}{MHz}—as employed by the WLAN 7 (802.11be) standard—already achieves an \ac{FNR} below \SI{10}{\%}. We also observe that the mean attack proves more effective in practice

\begin{figure}[htb]
   \centering
   \begin{subfigure}[b]{0.225\textwidth}
       \centering
       \includegraphics[width=1\linewidth]{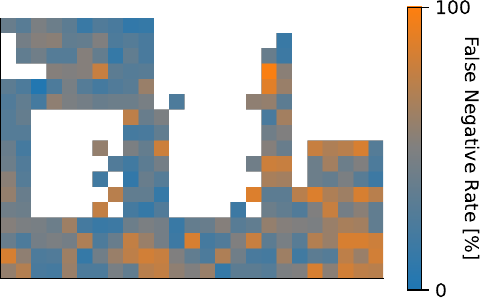}
       \caption{Bandwidth: \SI{20}{MHz}}
       \Description{todo}
       \label{}
   \end{subfigure}
   \quad
   \begin{subfigure}[b]{0.225\textwidth}
       \centering
       \includegraphics[width=1\linewidth]{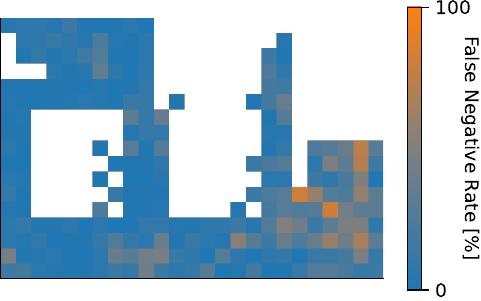}
       \caption{Bandwidth: \SI{160}{MHz}}
       \Description{todo}
       \label{fig:spatial_sec}
   \end{subfigure}
\caption{Spatial distribution of \ac{FNR} during the compensation attack against the \ac{RIS}-assisted \ac{ATR} with $f_c = $ \SI{5.31}{GHz}. (a) With \SI{20}{MHz} bandwidth and (b) with \SI{160}{MHz}.}
\end{figure}

Finally, we examine the spatial distribution of the \ac{FNR} in \autoref{fig:spatial_sec}. Each of the 250 drill positions is color-coded according to its \ac{FNR}, with the holes arranged two-dimensionally to mirror their physical location (cf. \autoref{fig:experimental_setup}). Note that the \ac{RIS} is located in the lower-left corner of these figures. At \SI{20}{MHz}, the distribution appears relatively uniform, whereas at \SI{160}{MHz} the \ac{FNR} rises more noticeably for positions that lie further along the plane from the \ac{RIS}. Especially the bottom right corner seems to be effected comparably less by the \ac{RIS}.

To conclude, we have demonstrated that the \ac{RIS}-assisted  \ac{ATR} allows to shift the root of trust from the infeasibility to exactly replicate the environment towards the secrecy of the \ac{RIS} configurations. 

\subsection{Adapting the Environment (C2)}

Another challenge of an \ac{ATR} system is usability optimization, referred to as challenge C2 in the Introduction. In particular, legitimate variation within the protected environment can cause false positive detections which are especially undesired in a tamper detection system. In the following, we consider a running ventilation fan as a legitimate moving component within the protected environment. We examine how the fan reduces the tamper detection performance and explore how the \ac{RIS} can mitigate this effect.

\paragraph{Effect of Threshold Adjustments} 
Crucially, the fan rotation causes significant temporal variation on the channel measurements, introducing a noise component. Thus, to prevent false positive detections, the \ac{ATR} threshold needs to be increased which in turn sacrifices the detection sensitivity -- a classical robustness-sensitivity trade-off. To evaluate this trade-off, we perform channel measurements for the untampered and tampered states at all drill positions for 50~random \ac{RIS} configurations. While we first measure with the fan being turned off, we repeat this measurement with the fan being turned on. \autoref{tab:fan_fp_fn} indicates the detection results in terms of the \ac{FPR} and \ac{FNR} for two threshold selection strategies. %

We first look at the column "Identical $\mathcal{T}$", where the threshold was selected as the maximum of the Euclidean distance when the fan is turned off and is not adjusted when turning the fan on. With the fan being off, this yields an satisfactory \ac{FNR} of \SI{23.83}{\%} without any false positives. However, when turning the fan on, the \ac{FPR} rises to \SI{83.35}{\%}. The result is that the initial balanced accuracy drops from \SI{88.09}{\%} to \SI{54.85}{\%}. Thus, to increase robustness, the system could be provisioned with the fan being turned on, employing a threshold that readily accounts for the fan being turned on. The result is indicated in the column "Separate $\mathcal{T}$". Here, the threshold is selected as the maximum of the Euclidean distance when the fan is turned on. While this reduces the \ac{FPR} again down to $0$, we now face a \ac{FNR} of \SI{76.98}{\%}. Thus, the tampering in many cases is not detected due to the threshold being chosen higher to accomodate the legitimate channel variation.

\begin{figure}[htb]
    \centering
    \includegraphics[width=0.75\linewidth]{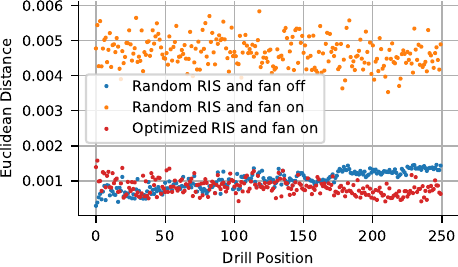}
    \caption{Untampered measurements under different conditions: RIS off with the fan on and off, and RIS on with the fan running.}
    \label{fig:intra_fan_compensation}
\end{figure}

\paragraph{\ac{RIS}-based Approach}
Taking advantage of the \ac{RIS}, we can adapt the smart radio propagation environment to mitigate the channel variation caused by the fan movement and thereby avoid the need for threshold adjustments. When provisioning the \ac{ATR}, we optimize the \ac{RIS} by means of the greedy optimization algorithm~\cite{tewesIRSenabledBreathTracking2022} while the fan is running. In particular, we formulate the following optimization goal, where we seek to find an  \ac{RIS} configuration $c_i$ that minimizes the average temporal standard deviation of $H(f, t_i, c_i)$:
\begin{equation}
    \min_{c_i} \text{avg}_{f} \left( \text{std}_t(H(f, t_i, c_i) )
    \right)
    \label{eq:opt_first}
\end{equation}
In our experiment, we optimize the \ac{RIS}, taking 20~consecutive measurements of $H(f, t_i, c_i)$, allowing us to use the standard deviation to assess the temporal fluctuation due to the fan movement. With the optimized \ac{RIS}, we successfully minimize the channel fluctuation due to the fan movement. This can be seen in \autoref{fig:intra_fan_compensation} where we plot the distances of channel measurements in the untampered state for the fan being turned off and on, for a randomized and an optimized \ac{RIS} configuration. Here, it becomes evident that the the Euclidean distance  significantly increases when comparing the fan being turned on and off. The important takeaway here is that the optimized \ac{RIS} configuration succeeds to reduce the Euclidean distance again down to the level of the fan being turned off, although the fan is still on. Thus, we conclude that the \ac{RIS} optimization successfully excludes the fan effect from the \ac{ATR}. Additionally to the previously discussed performance with randomly chosen \ac{RIS} configurations, \autoref{tab:fan_fp_fn} also reports the detection performance with the optimized \ac{RIS} configuration. Here, we can see that \ac{FPR} and \ac{FNR} values remain reasonably stable when turning the fan on, confirming that the \ac{RIS} optimization has successfully removed the effect of the fan whithout sacrificing the \ac{ATR} performance.

\begin{table}[h]
\footnotesize
\centering
\caption{Mean and standard deviation of detection performance of the optimized configuration against randomly generated configurations with activated and deactivated fan. We distinguish two cases: $(i)$ The \ac{ATR} is provisioned during a deactivated fan and the threshold $\mathcal{T}$ remains unaltered when the fan is turned on and $(ii)$ the threshold $\mathcal{T}$ is reset after the fan is turned on.}

\begin{tabular}{lrrrr}
\toprule
         $c_i$  &             Metric & \multicolumn{2}{c}{Identical $\mathcal{T}$}              & \multicolumn{1}{c}{Separate $\mathcal{T}$} \\ \midrule
\multicolumn{1}{l}{}             & \multicolumn{1}{c}{} & \multicolumn{1}{c}{Fan Off} & \multicolumn{1}{c}{Fan On} & \multicolumn{1}{c}{Fan On}         \\ \midrule
\multirow{3}{*}{Random}    & FPR {[}\%{]}         & 0.00 $\pm$ 0.00             & 83.35 $\pm$ 6.48           & 0.00 $\pm$ 0.00                            \\
                                 & FNR {[}\%{]}         & 23.83 $\pm$ 8.11            & 6.94 $\pm$  4.22          & 76.98 $\pm$ 8.22                           \\
                                 & ACC {[}\%{]}    & 88.09 $\pm$ 4.06            & 54.85 $\pm$ 2.11           & 61.51 $\pm$ 4.11                           \\ \midrule
\multirow{3}{*}{Optimized} & FPR {[}\%{]}         & \multicolumn{1}{c}{0.00}              & \multicolumn{1}{c}{2.40}            & \multicolumn{1}{c}{0.00}                            \\
                                 & FNR {[}\%{]}         & \multicolumn{1}{c}{24.80}             & \multicolumn{1}{c}{20.00}           & \multicolumn{1}{c}{22.80}                           \\
                                 & ACC {[}\%{]}    & \multicolumn{1}{c}{87.60}            & \multicolumn{1}{c}{88.80}            & \multicolumn{1}{c}{88.60} \\ \bottomrule                          

\end{tabular}
\label{tab:fan_fp_fn}
\end{table}

\begin{figure}[htb]
    \centering
    
    \begin{subfigure}[b]{0.225\textwidth}
        \centering
        \includegraphics[width=\linewidth]{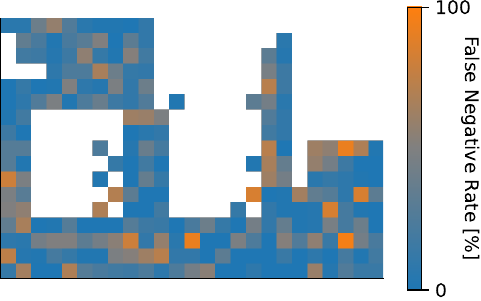}
        \caption{Random $c_i$ and fan off.}
        \label{fig:rand_fan_off}
    \end{subfigure}
    \quad
    \begin{subfigure}[b]{0.225\textwidth}
        \centering
        \includegraphics[width=\linewidth]{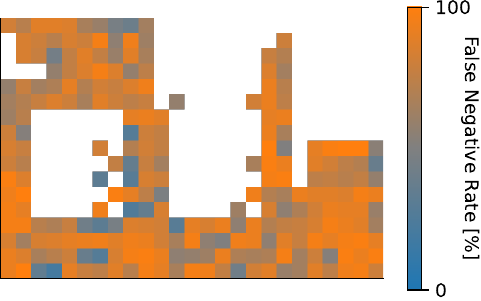}
        \caption{Random $c_i$ and fan on.}
        \label{fig:rand_fan_on}
    \end{subfigure}
    \quad
    \begin{subfigure}[b]{0.225\textwidth}
        \centering
        \includegraphics[width=\linewidth]{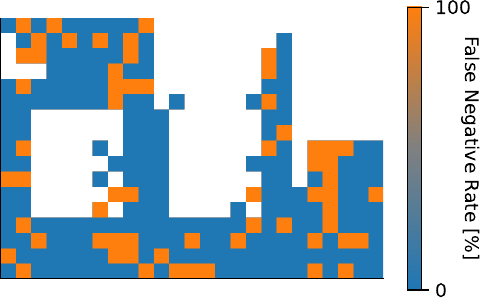}
        \caption{Optimized $c_i$ and fan off.}
    \end{subfigure}
    \quad
    \begin{subfigure}[b]{0.225\textwidth}
        \centering
        \includegraphics[width=\linewidth]{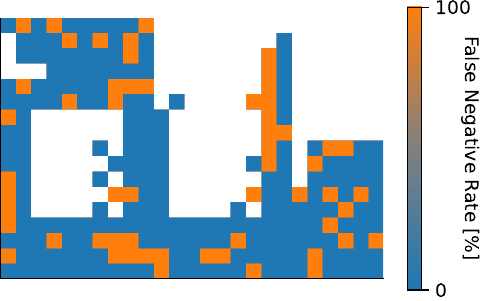}
        \caption{Optimized $c_i$ and fan on.}
    \end{subfigure}
    \quad
\caption{Local behaviour of average prediction for needle insertions with \SI{20}{MHz} bandwidth ($f_c$ = \SI{5.31}{GHz}). In (a) and (b) the baseline results with random configurations are shown. In (c) and (d) results for configurations optimized to minimize the fan effect are shown.}
\label{fig:fan_optim}
\end{figure}

\autoref{fig:fan_optim} further illustrates the spatial distribution of the fan's impact on the \ac{FNR} across different insertion positions, comparing both random and optimized configurations. When the fan is turned off there exist sensitive and insensitive needle positions. Not all positions are predicted correctly. With random $c_i$, the number of wrong predictions increases drastically when the fan is turned on. In the case of optimized $c_i$ the number of wrongly predicted insertions remains nearly unchanged, demonstrating that the fan exclusion mechanism effectively mitigates the fan’s influence on detection performance.

\paragraph{Does Fan Exclusion Degrade Spatial Sensitivity?}

One might ask whether the \ac{RIS} optimization excluding the fan variations from the channel response leads to a new attack vector, since we may exclude a spatial direction from the channel response. Our fan was located at the top center of the enclosure (cf. \autoref{fig:experimental_setup}, red fan). However, \autoref{fig:fan_optim} shows that there is no drastic change in the location of false negative events, countering the argument.

\section{Discussion and Future Work}

In this section, we discuss the experimental setup and our results. Moreover, we provide directions for future research.

\subsection{Experimental Setup and Results}

To facilitate the experimental evaluation of the underlying physical principles, our setup slightly deviates from a real-world \ac{ATR} deployment scenario. First, we adopted the evaluation methodology from~\cite{staatAntiTamperRadioSystemLevel2022}, using probing attack simulation as a prototypical attack scenario. To this end, we prepared the lid of the protected computer with pre-made holes to insert the needles. We placed the channel measurement system and the \ac{RIS} controller outside of the computing device. Under real-world conditions, the system would have to be completely self-contained. 

For the channel measurement system, we employed a~\ac{VNA}. While the \ac{VNA} is a rather expensive instrument, such measurements can likewise be obtained using low-cost wireless sensing solutions, \eg, based on \mbox{Wi-Fi}~\ac{CSI}~\cite{keerativorananMitigation2018} or \ac{UWB}~\cite{staatAntiTamperRadioSystemLevel2022}. However, the \ac{VNA} ensures high-quality results, thereby allowing to focus on the physical mechanisms at play. Moreover, the complex-valued \ac{VNA} measurements were a precondition to extract the attacker imprint and simulate the signal compensation attack. In a practical scenario, the attacker would need to synchronize their signal with the legitimate \ac{ATR} signals and pre-distort their signal to compensate the channel between the attacker and the \ac{ATR} receiver.

\paragraph{\ac{RIS} deployment}

In our experimental setup, it was easily possible to mechanically retrofit the prototypical \ac{RIS} into the computer. While the \ac{RIS} was initially designed for far-field applications, our experiments showed that it also is effective in the near-field scenario to reconfigure the \ac{ATR} radio environment. In an actual deployment scenario, additional engineering would allow adaption of the \ac{RIS} design to better fit the environment, both in terms of mechanical constraints and to optimize \ac{RF} performance, such as operating frequency and signal loss. The \ac{RIS} we used effectively impacts the radio signal propagation across a bandwidth of approximately~\SI{1}{\GHz} bandwidth, despite originally engineered to operate at a single frequency. Still, the limited bandwidth of the \ac{RIS} does not accomodate integration with a wideband \ac{ATR} system. Moreover, the \ac{RIS} optimization is most efficient in narrowband cases, \eg, when optimizing signal power for one particular frequency instead of a wide bandwidth. However, this conflicts with the \ac{ATR} being more likely to offer better detection rates in a larger bandwidth. Thus, if employing optimized \ac{RIS} configurations, the system designer needs to balance the \ac{ATR} bandwidth with the optimization efficiency. However, when employing the \ac{RIS} for mere security enhancement, the \ac{ATR} bandwidth can align with the \ac{RIS} bandwidth, as long as long as conforming with regulatory constraints. 

\subsection{Future Work}

In this work, we examined an \ac{RIS}-assisted \ac{ATR} system, showing that the \ac{RIS} allows to solve critical challenges of conventional \ac{ATR} systems. We believe that our work highlights the promising potential of combining the \ac{ATR} with the \ac{RIS} technology, paving the way towards further exploration.

\paragraph{Multi-Objective Utility}
As we have seen, the \ac{RIS} allows reducing the \ac{ATR} signal bandwidth while also enhancing robustness against noise sources in the environment. Moreover, we have shown that the \ac{RIS}-assisted \ac{ATR} can randomize the radio environment to withstand signal manipulation attacks. Future work should investigate how to utilize the \ac{RIS} multi-objectively to simultaneously realize the applications we have demonstrated. For example, the \ac{RIS} optimization could be repeated to obtain many different \ac{RIS} configurations that can be used at random.

\paragraph{Long-Term Stability}
Another interesting use-case of the \ac{RIS} would be to enhance the long-term system stability against environmental variation, \ie, to combat variation from legitimate environmental variation such as temperature, humidity, and aging. Here, the \ac{RIS} could be used to renew reference measurements in a secure fashion. We have additional work underway to investigate this particular application scenario.

\paragraph{Virtual Proofs of Reality}
One particularly interesting application that could be realized using \acp{RIS} are physical challenge-response protocols as proposed by Rührmair~\etal~\cite{ruehrmairVirtualProofsofReality2015} and Tobisch~\etal~\cite{tobischRemoteInspectionAdversarycontrolled2023}, allowing remote authentication and integrity verification based on querying channel measurements corresponding to random \ac{RIS} configurations. Future work should investigate whether such a system meets the requirement of hard-to-predict channel responses.

\section{Related Work}

In the following, we provide an overview of related research works on tamper detection. Compared to other areas of hardware security, there are relatively few contributions from the scientific community. One reason for this is the reliance on proprietary company~\ac{IP} and partially sensitive areas of application, making detailed descriptions of real-world anti-tamper measures, let alone attacks, scarce. Thus, the field is sometimes deemed to follow security-by-obscurity principles~\cite{helfmeierBreakingEnteringSilicon2013,  andersonSecurityEngineeringGuide2008, obermaierPresentFuturePhysical2018}. Notable exceptions are the detailed outlines of \ac{HSM} hardware security measures by Anderson~\cite{andersonSecurityEngineeringGuide2008}, Isaacs~\etal~\cite{isaacsTamperProofTamper2013}, and Obermaier and Immler~\cite{obermaierPresentFuturePhysical2018}.

Several research works study tamper detection mechanisms. Here, in the context of our paper, the most relevant work is the initial \ac{ATR} proposal by Staat~\etal~\cite{staatAntiTamperRadioSystemLevel2022} who demonstrated the general feasibility of radio-wave based tamper detection in a running 19"~server. We build on their work, likewise using automated probing attacks to evaluate an \ac{RIS}-assisted \ac{ATR}. A similar approach was later evaluated by Awal~\etal~\cite{awalNearfieldRFSensing2022}, likewise retrofitting antennas into a computer case to detect tampering. Xu~\etal employ impedance variations of bus lines on a \ac{PCB} to detect tampering such as probing~\cite{xuBusAuthenticationAntiProbing2020}. Impedance analysis of the power-delivery network of an \ac{FPGA} allows detection of tampering such as adding resistors or polishing the chip package, as demonstrated by Mosavirik~\etal~\cite{Mosavirik2023}. Immler~\etal~\cite{immlerSecurePhysicalEnclosures2019} proposed a protective cover to enclose a \ac{PCB}, acting as a~\ac{PUF} to derive an encryption key, thereby removing the requirement constant online self-monitoring. Another \ac{PUF}-based anti-tamper solution was put forward by Tuyls~\etal~\cite{tuylsReadProofHardwareProtective2006} to protect chips.

Several studies describe attacks against anti-tamper solutions. Anderson and Kuhn~\cite{andersonTamperResistanceCautionary1996} describe invasive attacks against tamper resistant chips. Using backside microprobing, Helfmeier~\etal~\cite{helfmeierBreakingEnteringSilicon2013} bypass the tamper detection of integrated circuits. Drimer~\etal~\cite{drimerThinkingBoxSystemLevel2008} used needle probing attacks to compromise the tamper detection of electronic payment systems. Weingart~\cite{weingartEvaluationSystemPhysical1990, weingartPhysicalSecurityDevices2000} presents efforts towards a systematic assessment of physical attacks and countermeasures.

\section{Conclusion}

In this work, we proposed the combination of an \ac{ATR} system with \ac{RIS} technology, thus combining radio-wave based physical tamper detection with smart radio environments. By placing an \ac{RIS} on the inside of a monitored computing system, the radio channel between the \ac{ATR} antennas -- the tamper detection modality itself -- becomes variable, offering a new degree of freedom for the design of an \ac{ATR} system. In particular, we have discussed key challenges of conventional \ac{ATR} systems that can be resolved in an \ac{RIS}-assisted \ac{ATR} system. Through extensive experimentation, we showed that the \ac{RIS} allows to shape the radio environment towards enhanced detection sensitivity, robustness, and security of the \ac{ATR} system. We believe that our work marks an important step towards resolving application-specific challenges of \ac{ATR} systems, overall fostering the acceptance of physical security solutions.

\begin{acks}
We thank Markus Heinrichs and Rainer Kronberger from the high-frequency laboratory at TH Köln for providing the \ac{RIS} prototypes. We thank Johannes Tobisch for helping with 3D printing. We thank Aaron Finkenthei for preparing the lid of the computer with the drill holes. This work was supported by the Deutsche Forschungsgemeinschaft (DFG, German Research Foundation) under Germany's Excellence Strategy - EXC 2092 CASA - 390781972.
\end{acks}

\bibliographystyle{ACM-Reference-Format}
\bibliography{atr_ris_bibliography}

\end{document}